\documentclass[aoas,preprint]{imsart}
\usepackage{comment,tabularx}
\usepackage{amsmath}

\usepackage{times}
\usepackage{bm}
\usepackage{amssymb}
\usepackage{graphicx}
\usepackage{natbib}
\usepackage[plain,noend]{algorithm2e}
\usepackage{caption}
\usepackage{breqn}
\usepackage{multirow}
\usepackage{xcolor}
\usepackage{multirow}
\usepackage{array}
\usepackage{tabularray}
\usepackage{diffcoeff}
\usepackage{amsthm}
\usepackage{appendix}
\usepackage{placeins}
\usepackage{url}

\usepackage{listings}
\usepackage{subcaption}

\def\bSig\mathbf{\Sigma}

\newcommand*{\rom}[1]{\expandafter\@slowromancap\romannumeral #1@}

\begin{document}
\begin{frontmatter}

\title{Stochastic Search Variable Selection for Bayesian Generalized Linear Mixed Effect Models}
\runtitle{SSVS for GLMM}

\begin{aug}
\author[A]{\fnms{Feng} \snm{Ding}},
\and
\author[B]{\fnms{Ian} \snm{Laga}\ead[label=e1]{ian.laga@montana.edu}}
\address[A]{Independent Researcher}
\address[B]{Department of Mathematical Sciences, Montana State University, Bozeman, MT, USA}
\runauthor{Ding and Laga}
\end{aug}

\label{firstpage}

\begin{abstract}
Variable selection remains a difficult problem, especially for generalized linear mixed models (GLMMs). While some frequentist approaches to simultaneously select joint fixed and random effects exist, primarily through the use of penalization, existing approaches for Bayesian GLMMs exist only for special cases, like that of logistic regression. In this work, we apply the Stochastic Search Variable Selection (SSVS) approach for the joint selection of fixed and random effects proposed in \cite{yang2020bayesian} for linear mixed models to Bayesian GLMMs. We show that while computational issues remain, SSVS serves as a feasible and effective approach to jointly select fixed and random effects. We demonstrate the effectiveness of the proposed methodology to both simulated and real data. Furthermore, we study the role hyperparameters play in the model selection.
\end{abstract}

\end{frontmatter}


\maketitle

\section{Introduction}
\label{sec:int}

Generalized linear mixed models (GLMMs) are a versatile statistical framework that extends the capabilities of the generalized linear model (GLM) by incorporating random effects to account for correlation and non-independence in the data. GLMMs are particularly useful when analyzing nested or clustered data, longitudinal studies, or any data structure where observations are not independent. A key challenge is determining which predictors to include in a GLMM model.

Methods for variable selection for linear mixed model (LMM) have been proposed over the years. Traditional methods such as backward/forward elimination by standard criteria such as the Akaike information criterion (AIC) or Bayesian information criterion (BIC) often fall short due to fact that the number of competing models grows exponentially with the number of predictors, creating significant computational challenges. In recent years, penalized regression with shrinkage estimation has emerged as an effective approach for selecting fixed effects \citep{Tiblasso, lassozou}. However, it has been pointed out that penalized methods tend to over-shrink large coefficients due to the relatively light tails of the Laplace priors \citep{yang2020bayesian}. Various frequentist tests have been developed to select random effects, including the global scope test \citep{lintest} and score test \cite{scoretest}. Bayesian approaches have also been developed to identify relevant random effects. For example, \cite{albertbaye} employed a mixture prior for the random effects variance and \cite{chen2003random} proposed a reparameterization method for random effects selection.

The joint selection of fixed and random effects remains a significant challenge with limited attention in the literature. This study aims to address this gap by extending existing Bayesian methods to GLMMs. A fully Bayesian approach for joint variable selection in logistic mixed-effects models was developed by \cite{Kinney2007}, building on the earlier work of \cite{chen2003random}. \cite{yang2020bayesian} later extended this idea, proposing a Bayesian shrinkage method to jointly select fixed and random effects specifically for linear mixed models (LMMs). However, to the best of our knowledge, no methods currently exist for jointly selecting fixed and random effects in more general GLMMs.

In this work, we show that the method proposed by \cite{yang2020bayesian} for LMMs can be directly applied to GLMMs. We demonstrate the effectiveness of this approach with both simulated and real data. Our objective is to identify significant subsets of fixed and random effects. We employ the Stochastic Search Variable Selection (SSVS) method, originally introduced by \cite{george1993variable}. The algorithm involves embedding the regression setting in a mixture model where latent variables are used to identify promising subsets of multiple regression covariates via Gibbs sampling. For random effects, we use the decomposition method of \cite{chen2003random} and reparameterize the GLMM to allow random effects to be selectively removed from the model. Shrinkage priors are then applied to these reparameterized terms to facilitate selection on the random effects. The algorithm is implemented in R using JAGS \citep{rlang, plummer2003jags, runjags}.

The structure of the paper is as follows. Section \ref{sec:methods} presents the GLMM framework, including reparameterization and prior specifications for the SSVS approach. Section \ref{sec:simu} presents a simulation study evaluating the feasibility and performance of our approach, while Section \ref{realdata} demonstrates its application using a real data example. We conclude with a discussion in Section \ref{conclusion}.
 
\section{Methodology}
\label{sec:methods}

In this section adapt the SSVS approach proposed in \cite{yang2020bayesian} to GLMMs. We summarize key details from their work for the reader's convenience.

\subsection{Generalized Linear Mixed Models}
Suppose there are $n$ subjects in a study and each subject has $n_i$ repeated observations for $i=1,...,n$. Let $Y_{ij}$ denote the response value for subject $i$ at measurement $j$. Let $X_{ij}$ be a $l \times 1$ design matrix for the fixed effects and $Z_{ij}$ be a $q \times 1$ design matrix for the random effects. Then the GLMM model is defined as:
\begin{align*}
    Y_{ij} | \mu_{ij} &\sim f(y|\mu_{ij}), \quad \mu_{ij} = E[Y_{ij}|q_i] \\
    g(\mu_{ij}) &= X_{ij}' \beta + Z_{ij}' \bm{\varrho}_i 
\end{align*}
where $f(y|\mu_{ij})$ is a known distribution from the exponential family, $\bm{\beta} = (\beta_1 ,..., \beta_l )^T$ is the vector of fixed effect coefficient, $\bm{\varrho}_i = (\varrho_{i1} ,..., \varrho_{iq} )^T \sim N(0,\bm{\Omega})$ is the $i$th random effect, and $g$ is the link function. $Z_{ij}$ is commonly chosen to be a subset of $X_{ij}$.  

As discussed in \cite{yang2020bayesian}, the covariance matrix can be decomposed into two matrices via the modified Cholesky decomposition proposed by \cite{chen2003random}, where
\begin{equation*}
    \bm{\Omega} = \bm{\Lambda} \bm{\Gamma} \bm{\Gamma}' \bm{\Lambda}',
\end{equation*}
where $\bm{\Lambda}$ is a diagonal matrix of the form $Diag(\lambda_1,...,\lambda_q),\, \lambda_k \geq 0,\, \forall k =1,...,q$, and 
\begin{equation*}
    \bm{\Gamma} = \begin{pmatrix}
    1 & 0 & \hdots & 0 \\
    \gamma_{21} & 1 & \hdots & 0 \\
    \vdots & \vdots & \vdots & \vdots \\
    \gamma_{q1} & \gamma_{q2}  & \hdots & 1
\end{pmatrix}
\end{equation*}
is a lower-triangular matrix with diagonal elements being 1 and each $\gamma_{uv}$ represents correlations between random effects. Note that $\bm{Lambda}$ is a diagonal matrix with elements proportional to standard deviations of the random effects, where zero-valued elements correspond to random effects having zero-valued variances, which is equivalent to removing the random effect from the model. This decomposition reparameterizes the covariance structure in a way that enables us to efficiently select random effects to include or exclude in the model.

Given this decomposition, the generalized linear mixed model can be expressed as 
$$
g(\mu_{ij}) = X_{ij}^' \beta + Z_{ij}^' \bm{\Lambda} \bm{\Gamma} \bm{\xi}_i
$$
where $\bm{\xi}_i = (\xi_{i1},...,\xi_{iq} )^'$ is a vector of independent standard normal latent variables. \cite{yang2020bayesian} pointed out that the elements of $\bm{\Lambda}$ and $\bm{\Gamma}$ are not independent; when $\lambda_k = 0$, the $k^{\text{th}}$ row and column of $\bm{\Gamma}$ are all set to $0$ except $\gamma_{kk} = 1 $. 

Now with the reparameterization, we can employ the SSVS algorithm to select influential random effects by putting a mixture prior with point mass at zero on each $\lambda_k$. When a $\lambda_k$ is set to be $0$ during the procedure, the constraints described above must be met.

\subsection{Prior Specification}
\label{sec:priors}
~\newline
\noindent \textit{Fixed Effects:} While \cite{yang2020bayesian} considered several different shrinkage priors, we consider only the generalized double Pareto prior. As discussed in \cite{yang2020bayesian}, we can select variables using the work of \cite{smith1996nonparametric} by introducing a latent variable $J_p, p = 1,...,l$, where $J_p=1 \text{ or }0$ indicates whether the $p^{\text{th}}$ effect is included in the model or not. Each $J_p$ is assigned an independent Bernoulli $1/2$ prior. The generalized double Pareto shrinkage priors are then given by
$$
\beta_{p: J_p = 1 } \sim N(0, \sigma^2 / g \theta_k ), \quad \theta_k \sim exp(\phi_k^2/2), \quad \phi_k \sim Gamma(1,1).
$$

~\\
\noindent \textit{Random Effects:} For each element $\lambda_k$ of $\Lambda$ we specify the following priors
$$
\lambda_k \overset{iid}{\sim} P_k \delta_0+(1-P_k)N_+(0,\tau_k^2h^2), \quad \tau_k^2 \sim IG(1/2,1/2),
$$
where each $P_k$ corresponds to a Bernoulli $1/2$ prior, $\delta_0$ denotes a point mass at zero and $N_+$ represents a truncated normal on the positive reals. The free elements of the lower triangular matrix $\Gamma$ are taken to be a vector $\bm{r} \in \mathbb{R}^ {q(q-1)/2 }$, which was assigned a multivariate normal prior 
$$
p(\bm{r}|\lambda) \sim N( \bm{\mu} , \bm{\Sigma} ) \text{ s.t. } \bm{r} \in R_\lambda.
$$
Here, $\bm{\mu}, \bm{\Sigma}$ are reasonable choices of the prior and $R_\lambda = \{ \bm{\gamma} : \gamma_{sk} = \gamma_{ks'} = 0 \text{ if } \lambda_k = 0, k=1,...,q, s = k+1,...,q, s' = 1,...,k-1 \}$, i.e., ``it constrains the elements of $\bm{r}$ to be zero when the corresponding random effects $\lambda_k$ are zero.''

For $\bm{\xi}_i$, \cite{yang2020bayesian} assign the generalized shrinkage prior
$$
\xi_{ik} \overset{ind}{\sim} N(0, \tau_k), \quad \tau_k \sim exp(m_k^2/2), \quad m_k \sim Gamma(1,1).
$$
The authors offered some insights on the prior choices of $\bm{\xi}_i$ regarding identifiability and computational efficiency, and we point the readers to their work for further information.

We implemented the above model in \texttt{runjags} via \texttt{R} \citep{plummer2003jags, runjags, rlang}.

\section{Simulations}
\label{sec:simu}

In this section, we evaluate the performance of the SSVS approach using simulated data. We take inspiration from the design of the simulation study by \cite{yang2020bayesian}, but modify the design to account for a Poisson GLMM and increase the number of possible variables to make variable selection more difficult. Specifically, we generate $N = 100$ data sets from the same distribution with $l = 10$ and $q = 10$, i.e.
\begin{align*}
    Y_ij &\sim \text{Poisson}(\mu_{ij}) \\
    \log(\mu_{ij}) &= X_{ij}' \beta + Z_{ij}' \bm{\varrho}_i,
\end{align*}
where
\begin{align*}
    \beta_1 &= 2 \text{ and } \beta_k \overset{iid}{\sim} Unif(-0.4, 0.4), &\text{for } k \in \{2, 3, 4, 5, 6 \}\\
    X_{ij} &\sim N(0, 1), &\text{for } j \neq 1\\
    \bm{\varrho}_{i} &\overset{iid}{\sim} N(\bm{0}, \bm{\Omega}), &\text{for } k \in \{1, 3, 6 \}
\end{align*}
\begin{equation*}
        \bm{\Omega} = 
\renewcommand{\arraystretch}{1} 
\setlength{\arraycolsep}{10pt} 
\begin{pmatrix}
0.08 & 0 & 0.04 & 0 & 0 & 0.02 & 0 & 0 & 0 & 0 \\
0 & 0 & 0 & 0 & 0 & 0 & 0 & 0 & 0 & 0 \\
0.04 & 0 & 0.15 & 0 & 0 & 0.09 & 0 & 0 & 0 & 0 \\
0 & 0 & 0 & 0 & 0 & 0 & 0 & 0 & 0 & 0 \\
0 & 0 & 0 & 0 & 0 & 0 & 0 & 0 & 0 & 0 \\
0.02 & 0 & 0.09 & 0 & 0 & 0.06 & 0 & 0 & 0 & 0 \\
0 & 0 & 0 & 0 & 0 & 0 & 0 & 0 & 0 & 0 \\
0 & 0 & 0 & 0 & 0 & 0 & 0 & 0 & 0 & 0 \\
0 & 0 & 0 & 0 & 0 & 0 & 0 & 0 & 0 & 0 \\
0 & 0 & 0 & 0 & 0 & 0 & 0 & 0 & 0 & 0
\end{pmatrix}
\end{equation*}

We simulate the fixed effects from a uniform distribution both so that the Poisson distribution was well behaved and so that we could evaluate whether a departure from the assumed normally distributed fixed effects would negatively impact the results.

We also consider two scenarios to study the performance of the variable selection approach under different setting. As in \cite{yang2020bayesian}, these two cases are designed to evaluate whether the variable selection approach is robust enough to ignore small, negligible effects. For Case 1, $\beta_k = 0$ for $k = 7, 8, 9, 10$, representing a sparse model. For Case 2, $\beta_k = 0.01$ for $k = 7, 8, 9, 10$, introducing small nonzero effects to evaluate robustness.

Lastly, we investigate the influence of the hyperparameters ($h$, $v$, and $\nu$) on model fit and parameter estimates. We fit all nine combinations of $h = 0.01, 1, 5$ and $v = \nu = 0.1, 1, 10$, where $v$ and $\nu$ vary together and are always equal.

In all cases, we fit three models: (1) the basic Poisson GLMM without shrinkage, (2) the SSVS Poisson GLMM with a diagonal matrix $\bm{\Gamma}$, and (3) the full SSVS Poisson GLMM with a lower triangular matrix $\bm{\Gamma}$. We refer to these models as the \textit{basic model}, the \textit{diagonal model}, and the \textit{full model}, respectively. Given the increased computational time associated with estimated a lower triangular matrix, we hope to see similar performance between the diagonal model and the full model.

\subsection{Case 1}

For each simulation and model fit, we calculated the selected model as the combination of fixed and random effects with the highest posterior density. Table \ref{tab:case1_topmodel} shows the percent of simulations which selected the most common models for the diagonal and full model for the hyperparameters $h = 1$ and $v = \nu = 0.01$. The true model was selected 44\% and 41\% of the time for the diagonal and full model, respectively. This result is consistent with our expectations, as sometimes the true coefficients are can be very small considering they are simulated from a uniform distribution between $-0.4$ and $0.4$. For almost all selected models, the random effect structure was correctly identified. Additionally, the root mean squared error (RMSE) is 0.000566 for the basic model, 0.000567 for the diagonal model, and 0.000539 for the full model.

Based on these results, we see that under this simulation design, the diagonal model performs similarly to the full model. However, for this design, the computation time is almost identical between the two models since the difference in the number of parameters is relatively small. We would expect to see a larger difference in computation time for a larger model, i.e., a model with more random effects.

\begin{table}[!t]
\caption{Percent of simulations in Case 1 where the considered model was selected with the highest posterior density for the diagonal model and the full model. Only the four more frequent models are included. The true model is in the first row.}
\label{tab:case1_topmodel}
\centering
\begin{tabular}{llll}
Model & & Diagonal & Full \\ \hline
$\beta_1, \beta_2, \beta_3, \beta_4, \beta_5, \beta_6 $ and &$\varrho_1, \varrho_3, \varrho_6$     &     44     &   41   \\
$\beta_1, \beta_2, \beta_3, \beta_4, \beta_5$ and &$\varrho_1, \varrho_3, \varrho_6$      &     7     &   9   \\
$\beta_1, \beta_2, \beta_3, \beta_4, \beta_5,\beta_6, \beta_8$ and &$\varrho_1, \varrho_3, \varrho_6$      &     6     &   5   \\
$\beta_1, \beta_2, \beta_3, \beta_4, \beta_5, \beta_6, \beta_7$ and &$\varrho_1, \varrho_3, \varrho_6$      &     4     &    5  
\end{tabular}%
\end{table}

\subsection{Case 2}

As for Case 1, Table \ref{tab:case2_topmodel} shows the percent of simulations which selected the most common models for the diagonal and full model for the Case 2 simulations. We see that compared to Case 2, it was much more likely for a model to select coefficients that were not in the true model. However, for the most frequently selected models, typically only one additional model was selected. While we do see a decrease in performance compared to the simulations with zero-valued coefficients, the proposed approach is relatively robust to small signals. Again, for almost all selected models, the random effect structure was correctly identified. The RMSE is 0.000587 for the basic model, 0.000560 for the diagonal model, and 0.000568 for the full model.

\begin{table}[!t]
\caption{Percent of simulations in Case 2 where the considered model was selected with the highest posterior density for the diagonal model and the full model. Only the four more frequent models are included. The true model is in the first row.}
\label{tab:case2_topmodel}
\centering
\begin{tabular}{llll}
Model & & Diagonal & Full \\ \hline
$\beta_1, \beta_2, \beta_3, \beta_4, \beta_5, \beta_6$ and &$\varrho_1, \varrho_3, \varrho_6$     &     21     &   24   \\
$\beta_1, \beta_2, \beta_4, \beta_5, \beta_6, \beta_8$ and &$\varrho_1, \varrho_3, \varrho_6$      &     8     &   8   \\
$\beta_1, \beta_2, \beta_3, \beta_4, \beta_5, \beta_6, \beta_9$ and &$\varrho_1, \varrho_3, \varrho_6$      &     8     &   7   \\
$\beta_1, \beta_2, \beta_3, \beta_4, \beta_5, \beta_6, \beta_7$ and &$\varrho_1, \varrho_3, \varrho_6$      &     6     &    6  
\end{tabular}%
\end{table}

\subsection{Hyperparameters}

Here we evaluate the influence of the hyperparameters on the full model by studying the percent of simulations which selected the true model and the RMSE of the fixed effects across the range of hyperparameters. These results are shown in Tables \ref{tab:hyper_topmodel} and suggest that the models are sensitive to very small values of $v = \nu$, but are relatively robust to varying values of $h$. These set of simulations suggest the optimal model hyperparameters are $v = \nu = 5$ and $h = 0.1$, but these results should not be assumed to hold for different data sets and designs.

It is noteworthy that even though the percent of simulations where the true model was recovered varies substantially for the different hyperparameter combinations, for Case 2, the RMSE is relatively stable. This suggests that even when the true model was not found, the selected model was very similar to the true model. This is consistent with our expectations since Case 2 technically involves true but negligible effects, so the model appears to occasionally select these negligible coefficients. On the other hand, the RMSE for the Case 1 simulations increases substantially for two of the hyperparameter coefficients. Based on the results from Table \ref{tab:case1_topmodel}, this may be the result of excluding true coefficients rather than including false coefficients.

\begin{table}[!t]
\caption{Percent of simulations for Case 1 and 2 where the true model was selected for each combination of hyperparameters are shown in the model percent columns, while the root mean squared error is shown in the RMSE columns.}
\label{tab:hyper_topmodel}
\centering
\begin{tabular}{llllll}
$v = \nu$ & $h$ & Case 1 Model Percent & Case 1 RMSE & Case 2 Model Percent  & Case 2 RMSE\\ \hline
0.01  & 0.1 &  26   & 0.001056 &  9  & 0.058032 \\
1  &   0.1 &   41 & 0.000539 &  24  &  0.055842\\
5  & 0.1 &  49& 0.000560  &  32   & 0.055743\\
0.01  & 1&   24 &  0.001456 &  10   & 0.056433\\
1  &  1&  42 &  0.000567 &  26&   0.055892 \\
5  & 1& 46&0.000542 &  32 &  0.055948 \\
0.01  &   10 & 17& 0.000553  &  5&   0.056864 \\
1  &  10&  33&  0.000565  &  20&  0.055953  \\
5  & 10&  34& 0.000536 &  22&    0.055863
\end{tabular}%
\end{table}

\FloatBarrier
\section{Application to Real Data}
\label{realdata}

In this section, we demonstrate the variable selection technique using a complex data set analyzed by \cite{rolhauser2020accounting} to study the trait-environment relationship in Wisconsin forests. The data set contains $34,652$ observations on 189 sites (site) and 185 species (sp). There is one observations for each site and species interaction. The authors fit the following Negative Binomial GLMM with an offset using the \texttt{glmmTMB} function in \texttt{R} \citep{glmmTMB, rlang}:
\begin{lstlisting}
glmmTMB(abundance~(MAP + MAT + TSD + %Sand + %N + BA
    + I(MAP^2) + I(MAT^2) + I(TSD^2)
    + I(%Sand^2) + I(%N^2)
    + I(BA^2)) * (VH + LS + LMA + LCC)
    + I(VH^2) + I(LS^2 + I(LMA^2) + I(LCC^2))
    + (1 + VH + LS + LMA + LCC | site)
    + (1 + MAP + MAT + TSD + %Sand + %N + BA | sp)
    + offset(log(quadrats))
\end{lstlisting}

Based on their final model and the available data, we implemented the proposed variable selection technique for a slightly more complex model by adding \texttt{PCV} as a fixed effect and a random slope for sp, i.e., the equivalent \texttt{glmmTMB} code ais given by
\begin{lstlisting}
glmmTMB(abundance~(MAP + MAT + TSD + PCV + %Sand + %X.N + BA
    + I(MAP^2) + I(MAT^2) + I(TSD^2)
    + I(%Sand^2) + I(%N^2)
    + I(BA^2) * (VH + LS + LMA + LCC)
    + I(VH^2) + I(LS^2) + I(LMA^2) + I(LCC^2)
    + (1 + VH + LS + LMA + LCC | site)
    + (1 + MAT + MAP + TSD + PCV + %Sand + %N + BA | sp)
    + offset(log(quadrats))
\end{lstlisting}

Note that as opposed to the simulation study, this model contains two independent random effect structures that are shrunk independently. For this analysis, we set $h = 1$ and $v = \nu = 0.01$, although based on the simulation results, the choice of hyperparameter should not substantially affect the parameter estimates. We ran the MCMC across 3 independent chains using 1000 adaptation samples, 1000 burn-in samples, and 3000 post burn-in samples, resulting in 9000 posterior samples used for inference.

The results indicate that no single combination of fixed effects substantially outperforms others. In fact, the highest posterior density of fixed effects considered independently of any random effect combination occurred in only 2/9000 posterior samples. The coefficient with the fewest nonzero posterior samples corresponded to \texttt{I(\%Sand^2)}, but was still nonzero in 4349/9000 of the posterior samples. We conclude that there are a large number of models which result in similar performance.

On the other hand, the top selected random effects are clearer. The model with only a random intercept and slopes for MAT and TSD with respect to the species groups was selected in 5451/9000 of the posterior samples. This model did not include any random components with respect to the sites. The second most common model, which included a random intercept for sites, was selected in 635 posterior samples.

\begin{figure}[!htp]
    \centering
    \begin{subfigure}{0.49\textwidth}
        \includegraphics[width=\textwidth]{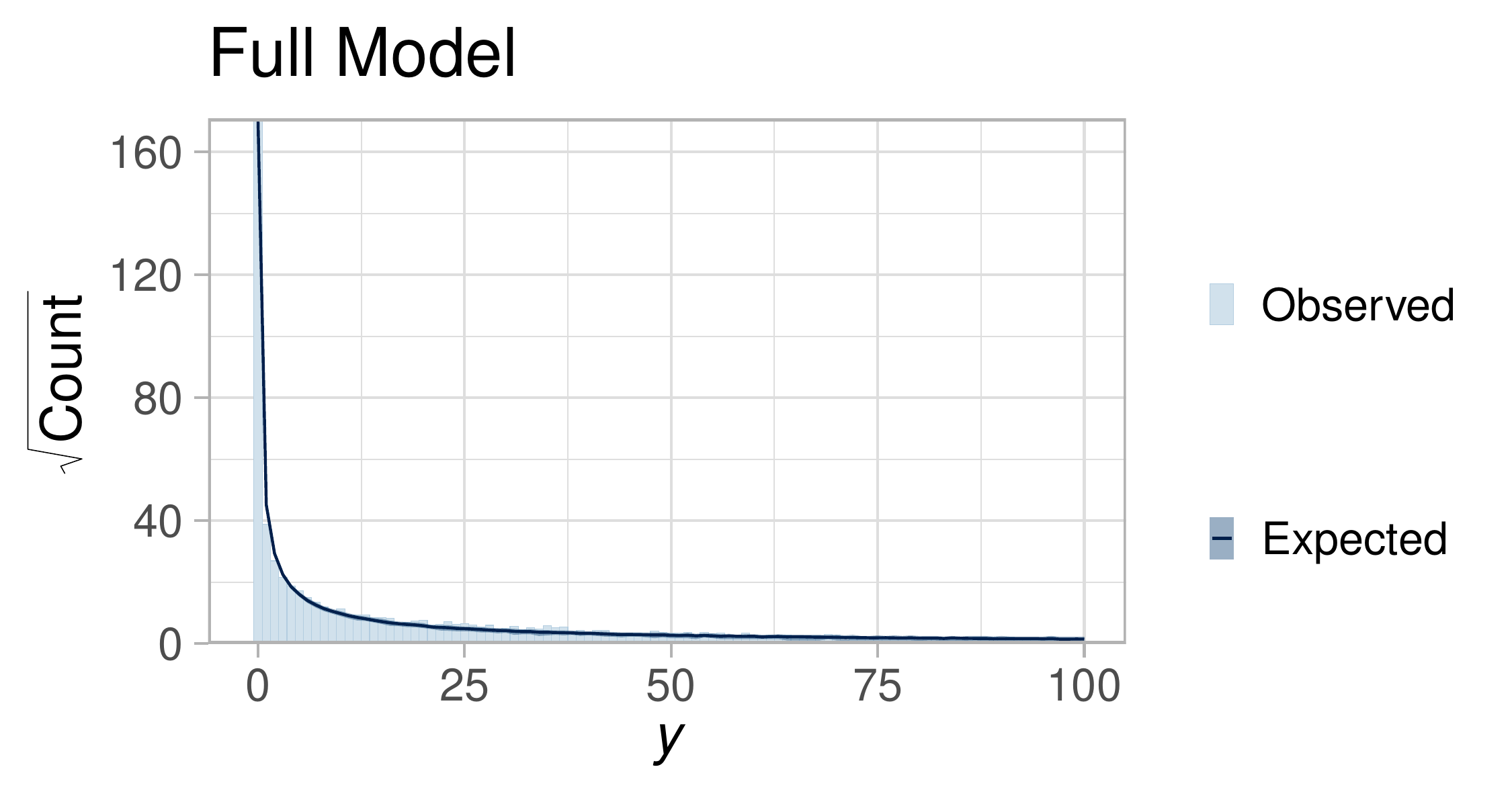}
      \caption{}
        \label{fig:full_root}
    \end{subfigure}
    \begin{subfigure}{0.49\textwidth}
        \includegraphics[width=\textwidth]{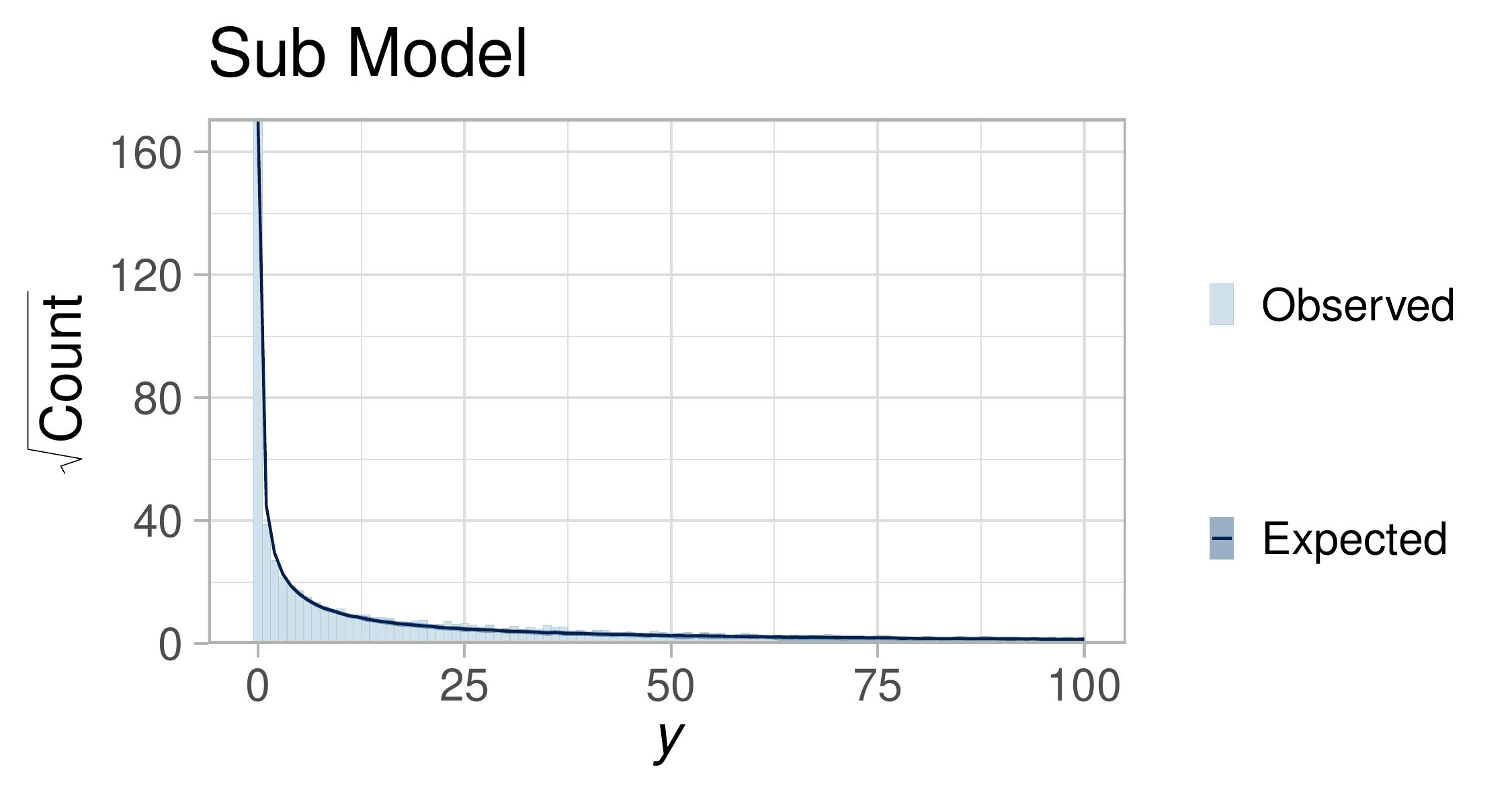}
      \caption{}
        \label{fig:sub_root}
    \end{subfigure}
    \begin{subfigure}{0.49\textwidth}
        \includegraphics[width=\textwidth]{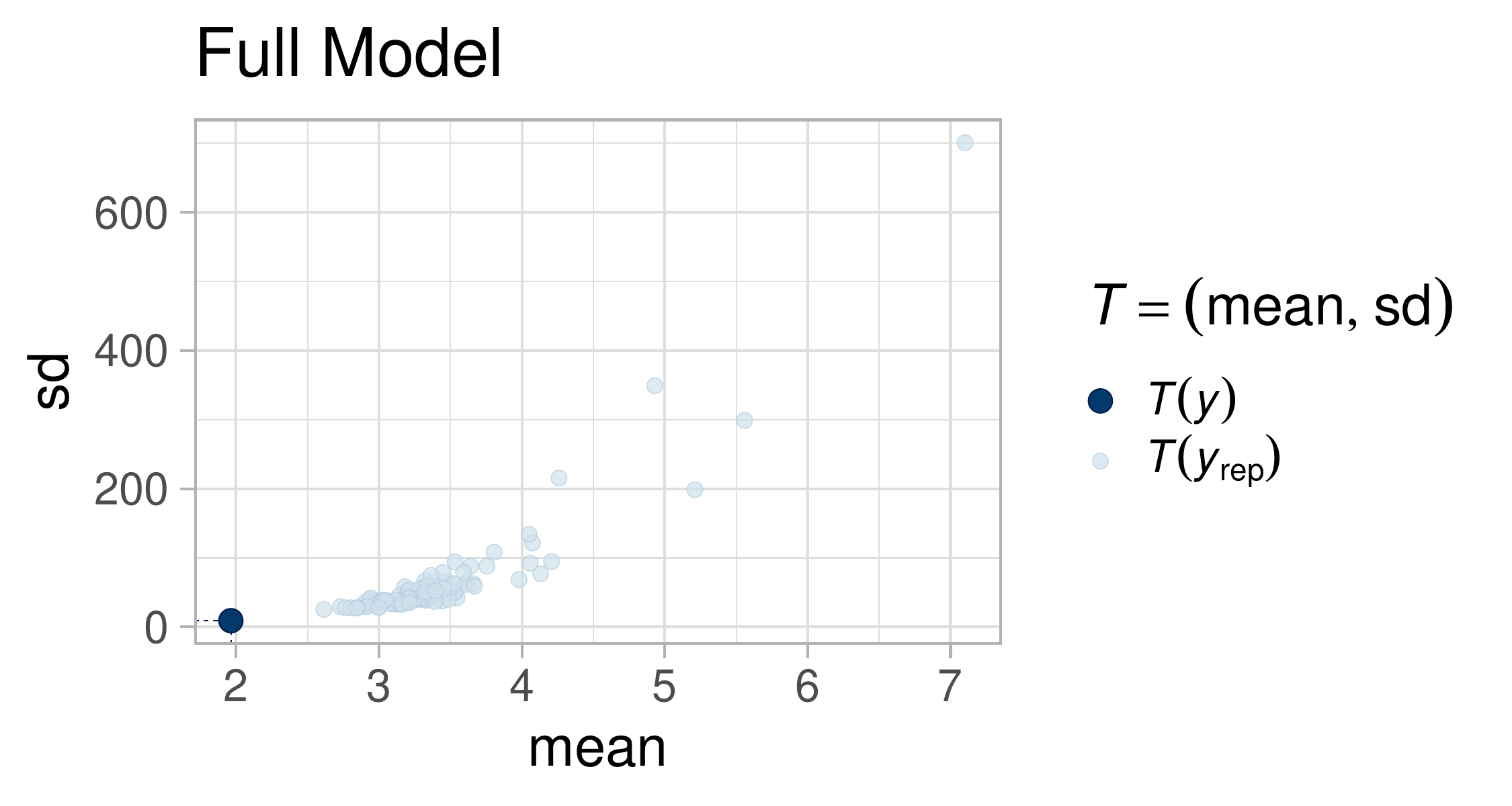}        
        \caption{}
        \label{fig:full_stat}
    \end{subfigure}
    \begin{subfigure}{0.49\textwidth}
        \includegraphics[width=\textwidth]{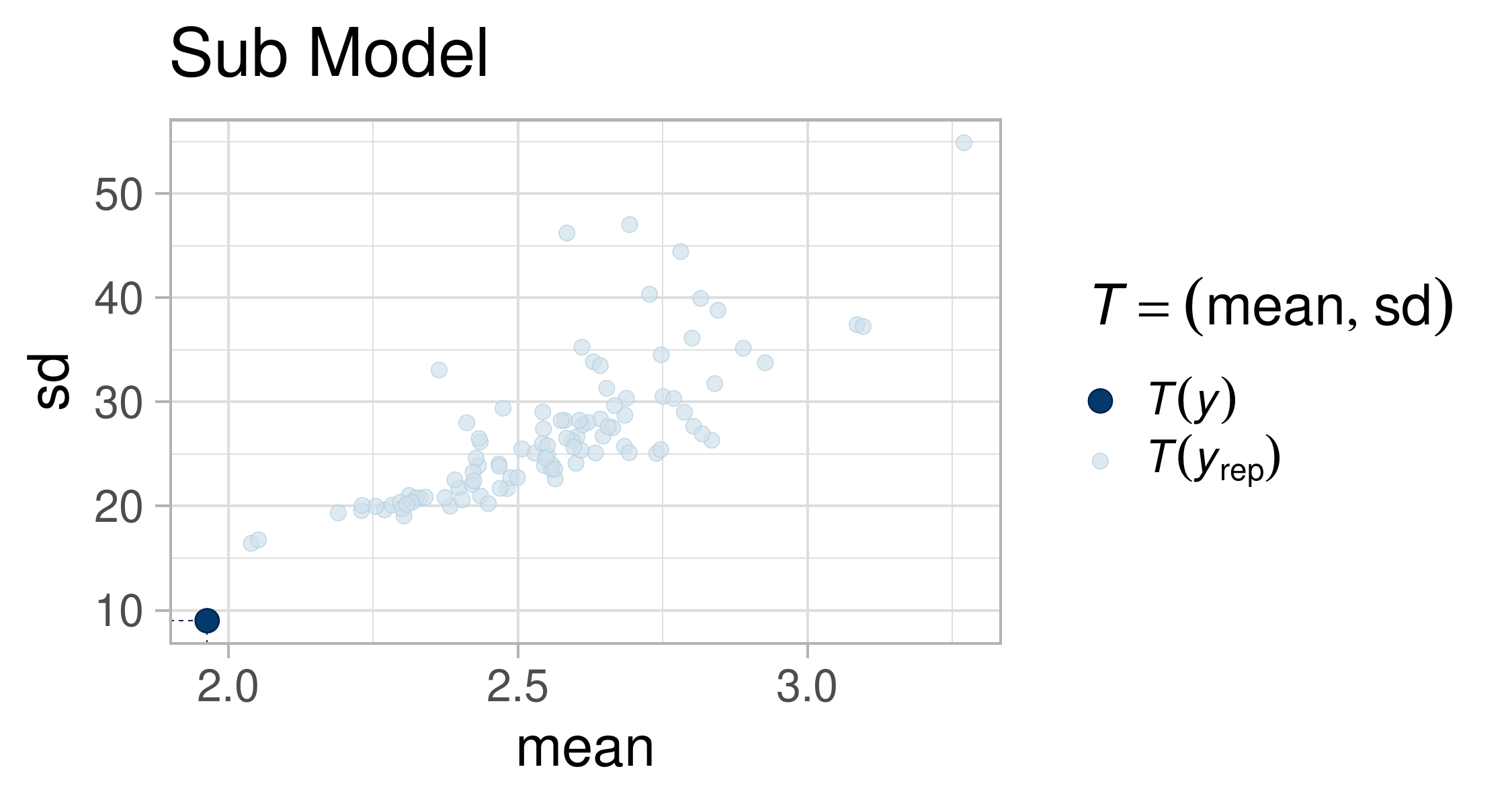}
       \caption{}
        \label{fig:sub_stat}
    \end{subfigure}
\caption{Posterior predictive checks for the full and sub model fit to the real data. Rootograms are shown for the full and sub model in (a) and (b), respectively, while a scatterplot of the mean and standard deviation are shown for the full and sub model in (c) and (d).}
\label{fig_sim2}
\end{figure}

The results from our variable selection approach differ dramatically from the considered model in \cite{rolhauser2020accounting}. To evaluate whether the variable selection is working appropriately, we studied diagnostic plots of a model refit with only the selected random effect structure. We refit the two models using a Bayesian GLMM without shrinkage via the \texttt{brms} package \citep{brms1, brms2}. We compared the model fits using posterior predictive checks. Although not all posterior predictive checks are included in this manuscript, we include the results from the rootograms (Figures \ref{fig:full_root} and \ref{fig:sub_root}) and scatterplots of the mean and standard deviation of the data (Figures \ref{fig:full_stat} and \ref{fig:sub_stat}). The rootograms suggest that the full model and the sub model produce nearly identical probability mass functions. On the other hand, the scatterplots show that the mean and standard deviation of posterior replicated data from the sub model is actually closer to the truth, suggesting that the sub model actually outperforms the larger model fit by the authors in their original manuscript.

While these results are not conclusive, they suggest that the proposed variable selection technique is able to accurately select a simpler model which performs just as well if not better as the more complicated full model. We did not further investigate the performance of the fixed effect structure, but the model indicates that any number of combinations of the fixed effects would result in similarly performing models.

\section{Conclusion}
\label{conclusion}

This study presents a novel application of the Stochastic Search Variable Selection methodology to Generalized Linear Mixed Models, filling a gap in the existing literature. We show that the results from previous work, especially the work \cite{yang2020bayesian}, can be extended directly to GLMMs with only a modification of the likelihood. The proposed methodology addresses the challenges posed by the high dimensionality and complex interactions frequently faced with fitting GLMMs. The results indicate that the approach is able to effectively remove small coefficients while retaining coefficients critical to explaining the underlying relationships in the data.

Simulation studies demonstrated the efficacy of the proposed method under various scenarios, including settings with both strong and weak signal-to-noise ratios. The results highlighted the ability of the SSVS approach to reliably identify true fixed and random effects, even in challenging conditions with small effect sizes or increased dimensionality. Furthermore, we found that the choice of hyperparameters did not play a significant role in the performance of the model of the selection of the coefficients, making the method easier and more robust to use in practice.

We also applied the proposed methodology to a real ecological data set with reasonable success. While the approach was relatively unsatisfying for this data set in selecting a unique set of fixed effects, the model was able to find a substantially simpler model with respect to the random effects. In fact, this simpler model seemed to outperform the full model with respect to some posterior predictive checks.

In summary, this work shows that researchers can apply existing variable selection approaches directly to GLMMs. We identify two remaining directions of research. First, we considered only the generalized Pareto prior for the fixed effects, so it is unclear whether other shrinkage priors would perform better in practice, especially with respect to the ecological data set we studied here.

Second, and perhaps more importantly, computation time remains a limiting factor. While the SSVS approach requires fitting only one or two models compared to the potential dozen models that a step-by-step variable selection approach would require, fitting the model with shrinkage priors is often overly restrictive. For our real data analysis, convergence was slower and we were unable to fit the full model in a reasonable amount of time. We hope that future researchers can develop a substantially faster variable selection approach for GLMMs, which would undoubtedly be useful to the broader research community.

\bibliographystyle{apalike}
\bibliography{refs}

\end{document}